\documentclass[a4paper,fleqn]{cas-dc}

\usepackage[numbers]{natbib}

\def\tsc#1{\csdef{#1}{\textsc{\lowercase{#1}}\xspace}}
\tsc{WGM}
\tsc{QE}
\tsc{EP}
\tsc{PMS}
\tsc{BEC}
\tsc{DE}
\usepackage{amsmath,amssymb,amsfonts}
\usepackage{algorithmic}
\usepackage{graphicx}
\usepackage{textcomp}
\usepackage{xcolor}
\usepackage{fixltx2e}
\usepackage{subfloat}
\usepackage{subfigure}
\usepackage{url}
\usepackage{booktabs}

\def\BibTeX{{\rm B\kern-.05em{\sc i\kern-.025em b}\kern-.08em
    T\kern-.1667em\lower.7ex\hbox{E}\kern-.125emX}}

\begin{document}

\let\WriteBookmarks\relax
\def\floatpagepagefraction{1}
\def\textpagefraction{.001}

\title[mode = title]{Ultra8T: A Sub-Threshold 8T SRAM with Leakage Detection\\
}
\tnotetext[1]{This work was supported by This work is supported by the National Natural Science Foundation of China (grant number 62204141, 62090025) and partially support by Tsinghua University Initiative Scientific Research Program.}

\author[1]{Shan Shen}
\ead{shanshen@tsinghua.edu.cn}

\author[2]{Hao Xu}

\author[3]{Yongliang Zhou}

\author[2]{Ming Ling}\cormark[1]
\ead{trio@seu.edu.cn}

\author[1]{Wenjian Yu}
\cormark[1]
\ead{yu-wj@tsinghua.edu.cn}

% \credit{Conceptualization of this study, Methodology, Software}

\address[1]{Dept. Computer Science \& Tech., BNRist, Tsinghua University, Beijing, China}
\address[2]{Nation ASIC System Engineering Technology Research Center, Southeast University, Nanjing, 210096, China.}
\address[3]{School of Integrated Circuits, Anhui University, Hefei
 230601, China.}
\cortext[cor1]{Corresponding author.}


\maketitle

\begin{abstract}
In energy-constrained scenarios such as IoT applications, the primary requirement for System-on-Chips (SoCs) is to increase battery life. However, when performing sub/near-threshold operations, the relatively large leakage current hinders Static Random Access Memory (SRAM) from normal read/write functionalities at the lowest possible voltage (V\textsubscript{DDMIN}). In this work, we first put forward a model that describes a specific relationship between read current and leakage
noise in a given column. Based on the model, Ultra8T SRAM is designed to aggressively reduce V\textsubscript{DDMIN} by using a leakage detection strategy where the safety sensing time on bitlines is quantified without any additional hardware overhead. We validate the proposed Ultra8T using a 256×64 array in 28nm CMOS technology. Post-simulations show successful read operation at 0.25V with 1.11\textmu s read delay, and the minimum energy required is 1.69pJ at 0.4V
\end{abstract}

\section{Introduction}
Wireless sensors designed for Internet-of-Things (IoT) applications have emerged as a prominent computing class in recent times and are typically severely power-constrained. As a result, the power budget of wireless sensors has become more restricted due to the shrinking battery size. Long battery life is essential for wireless sensors in many IoT applications, leading to the requirement for energy-efficient operations \cite{lee2019self}. One of the most direct and effective methods is to aggressively scale the supply voltage (V\textsubscript{DD})
to the near-/sub-threshold regions, as the active power has a quadratic dependence on voltage and the leakage power is exponentially related to the voltage. However, reducing voltage can compromise the robustness of the circuits and even cause the systems to malfunction. Therefore, maintaining a low minimum operating voltage (V\textsubscript{DDMIN}) to achieve the highest energy efficiency while keeping the circuits robust enough is of paramount importance for these power-constrained systems.

Static Random Access Memory (SRAM) is one of the most crucial components of low-power systems and continues to be a bottleneck to lower V\textsubscript{DD} minimum \cite{alioto2017enabling}. As the supply voltage decreases, V\textsubscript{DDMIN} of the SRAM module has a strong correlation with the variation of the threshold voltage (V\textsubscript{TH}) and is susceptible to parameter mismatch of the devices. In the conventional 6T SRAM, V\textsubscript{DDMIN} is limited due to conflicting requirements of read stability and write ability. In addition, read-disturb and half-select destruction further deteriorate the robustness of the 6T cell at low supply voltages. Consequently, the 6T SRAM needs to adopt various read and write assist techniques at near-threshold voltages (0.5V~0.7V in this work) to ensure reliable access \cite{chang2013sub}. 
Other cell structures are proposed to provide good read/write stability by adding more transistors to separate the read and write ports. For example, 8T SRAM is widely studied in several low-power systems \cite{chang20088t,yu202265}, but still faces challenges when operating at sub-threshold voltages ($<$ 0.5V in this work).

During a read operation, the swing of the Bit-Line (BL) is caused by both the read current from the accessed cell and the total leakage noise from all idle cells in the same column. As V\textsubscript{DD} scales down, the on-and-off current ratio from transistors becomes smaller, making the on-current more susceptible to the Process Voltage Temperature (PVT) condition. This implies that the BL swing caused by leakage noise increases, reducing the contribution of the read current from the accessed cell. Additionally, since the cutoff current is greater through a `0' cell than through a `1' cell, the data pattern in the accessed column and the column depth affect the total leakage current. Consequently, in the sub-threshold domain, a Sense Amplifier (SA) cannot differentiate between a BL swing caused by reading a `0' cell with the minimum leakage or reading a `1' cell with the maximum leakage.

Several techniques have been proposed to enhance the resilience of SRAM at ultra-low voltages. Novel cell structures \cite{do2012sensing, chien20180} have been proposed to suppress the interference of cell leakage current on RBL, but this inevitably comes with increased area overhead. Read/write assistant circuits \cite{verma2008256,do201532kb, wen2016bit, do20160, do2019energy} reduce V\textsubscript{DDMIN} by avoiding worst-case scenarios and adopting PVT-resistant structures, but they either introduce additional hardware overhead or increase the access delay. Timing speculations \cite{shen2020modeling, ling2021design} have been employed as other strategies, but they cannot further reduce the V\textsubscript{DDMIN} since the error correction overhead becomes too large in the sub-threshold domain.

In this study, we present Ultra8T SRAM with a leakage detection strategy that enables robust read operations from the nominal voltage (0.8V) down to the sub-threshold voltage (0.25V). Compared to other sub-threshold SRAMs, the proposed SRAM achieves the best access delay 1.1\textmu s at 0.25V. Our core idea is based on a model that describes a specific relationship between read current and leakage current in a given column. By measuring the leakage current in advance, a safety sensing window can be defined. The model is implemented in simple hardware where the leakage current is first converted into digital pulses by SA circuits, then, the read timing is configured according to the quantified values. Moreover, during normal read access, there is no additional power or delay overhead in SRAM.
%Since the leakage noise is pre-measured, Ultra8T can be operated at an aggressive V\textsubscript{DDMIN}. 
The aforementioned strategy is enhanced by two critical circuitries: (i) a digitized timing scheme that generates PVT-tracking and variance-suppressed internal clock and (ii) a sensing amplifier with high sensitivity \cite{sosa}. 

The rest of this paper consists of 5 sections. Section II discusses the issues and existing techniques related to low-power SRAMs. Section III presents the design of Ultra8T SRAM. Section IV provides simulation results and comparisons. Finally, in Section V, we conclude the whole paper.

\section{Backgrounds}

\subsection{8T SRAM Read Access}

\begin{figure}[tb]
\centering
\includegraphics[width=0.8\linewidth]{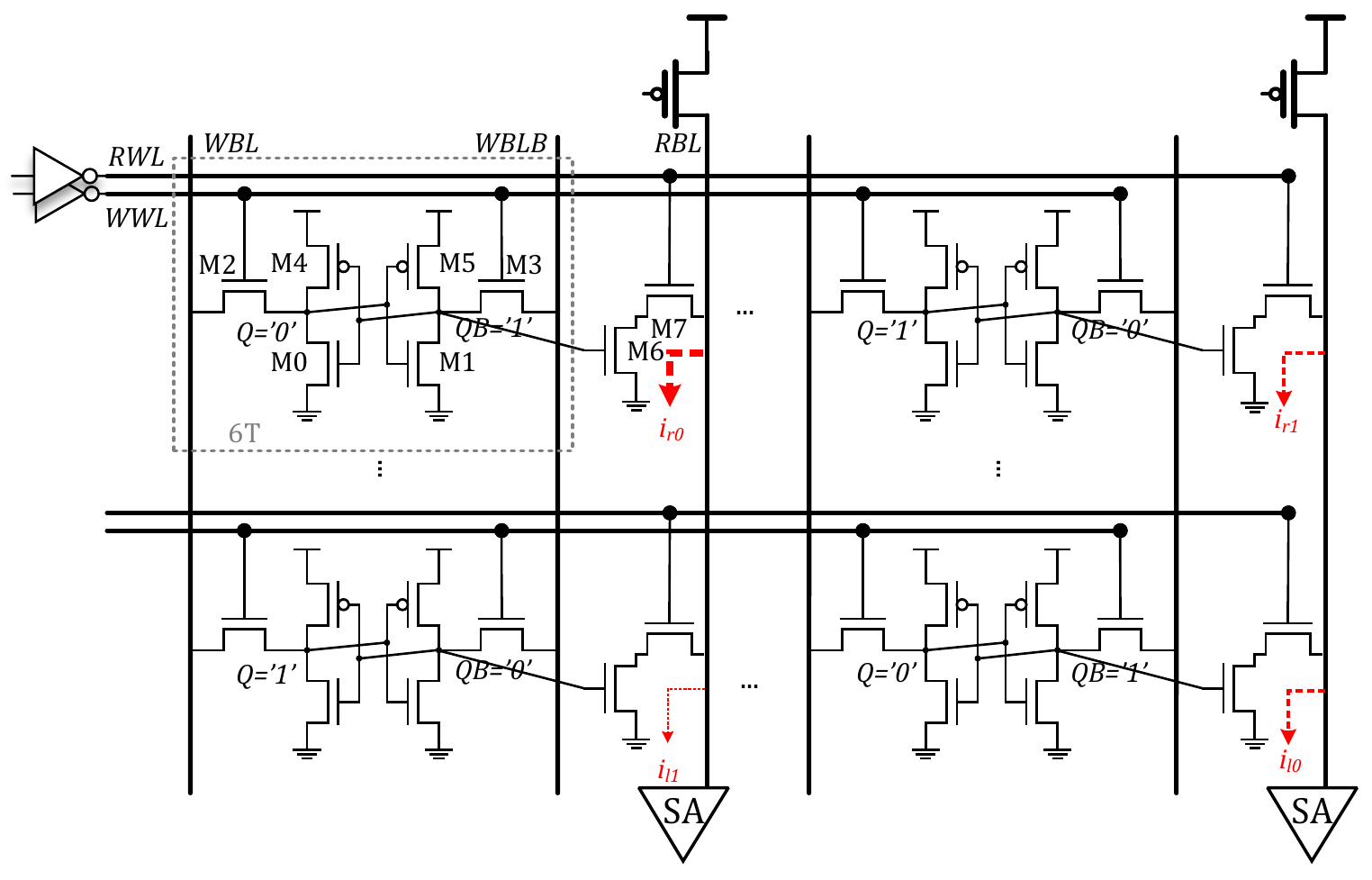} 
\caption{Schematic of 8T SRAM column.}\label{fig1}
\vspace{-0.4cm}
\end{figure}

Figure \ref{fig1} depicts a typical 8T SRAM organization. Each row of SRAM contains a Read Word Line (RWL) that connects to read ports and a Write Word-Line (WWL) that connects to the write ports of SRAM cells. Each column has a pair of complementary Write Bit-Lines (WBL and WBLB) for writing into cells and a single Read Bit-Line (RBL) for reading. The write operation for 8T SRAM is the same as that of conventional 6T SRAM, while the read operation is performed through the read port. During standby, precharging PMOS devices precharge all RBLs to V\textsubscript{DD}. When the read cycle is activated, the PMOSs are turned off, leaving RBLs floating. Ideally, the RBL is then discharged to the ground if the accessed cell stores a `0' and stays at V\textsubscript{DD} otherwise. 

\begin{figure}[tb]
\centering
\subfigure[]{\includegraphics[width=0.8\linewidth]{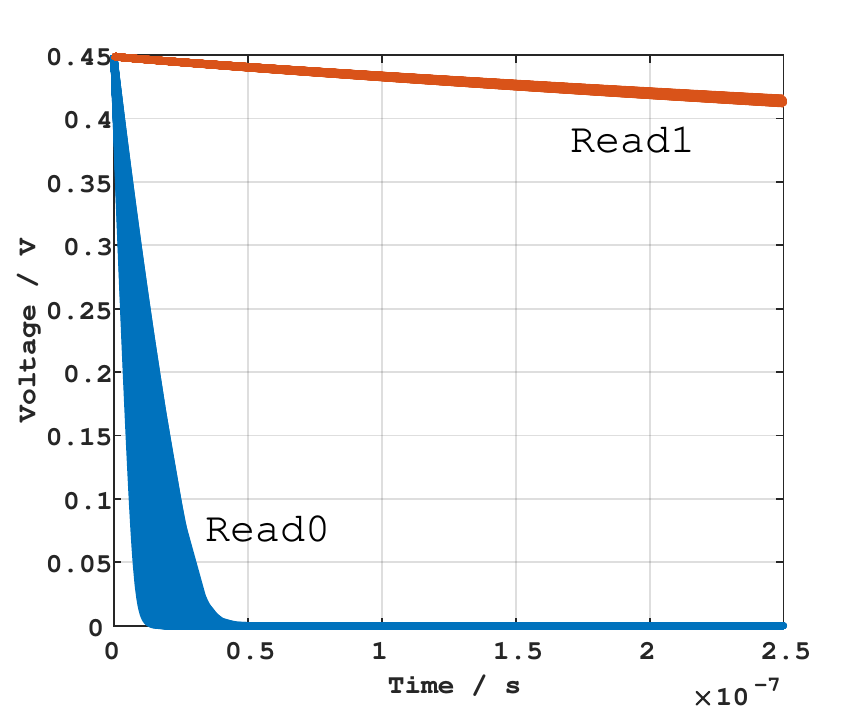} \label{fig2-a}}
\subfigure[]{\includegraphics[width=0.8\linewidth]{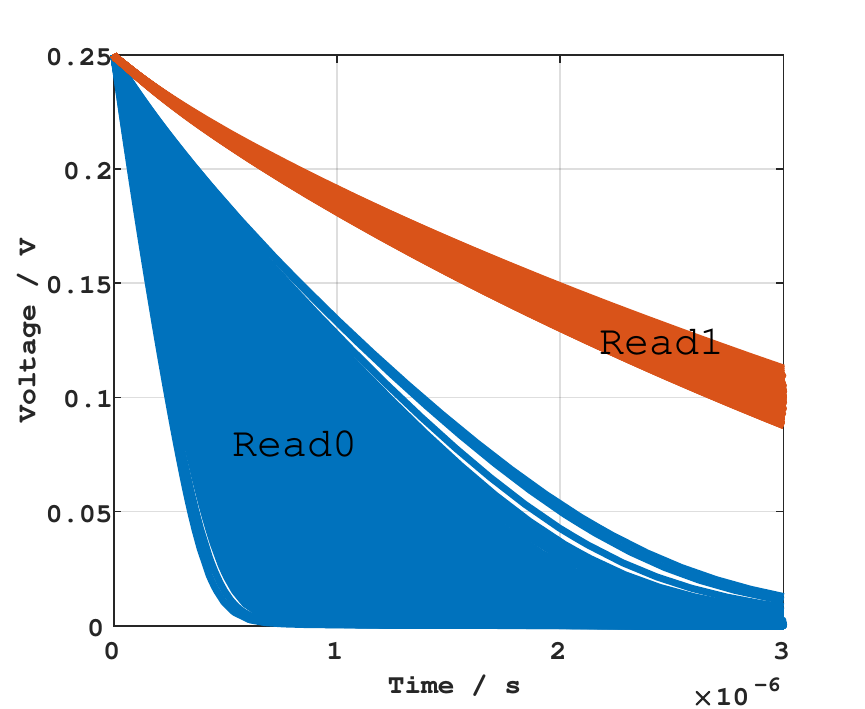} \label{fig2-b}}
\caption{The worst-case for reading ‘1’ and ‘0’ in a 256-rowed column at (a) 0.45V and (b) 0.5V V\textsubscript{DD}.}\label{fig2}
\vspace{-0.4cm}
\end{figure}

As illustrated in Fig. \ref{fig2}, there are 2 types of current associated with the RBLs: read current $i_{r}$ for reading `0' ($i_{r0}$) and `1' ($i_{r1}$), $i_l$ is the total leakage comprised of current leaking through `1' idle cells ($i_{l1}$) and `0's ($i_{l0}$). Ideally, both $i_{l0}$ and $i_{l1}$ are close to zero, $i_{r1} \approx i_{l0} \approx i_{l1} \ll i_{r0}$. Thus, the sensing circuit can differentiate easily between $i_{r1}$ and $i_{r0}$. However, leakage current becomes non-negligible under low V\textsubscript{DD} or when a large number of cells connect to one RBL because the on-current magnitude becomes comparable to that of cutoff current as the voltage supply scales down. Moreover, $i_{l}$ is dependent on data stored in SRAM cells. In Fig. \ref{fig1}, $i_{l1}$ flows through two stacked off-transistors while $i_{l0}$ flows through a single off-transistor in the leakage path, which makes $i_{l0} < i_{l1}$. At low voltages, the accumulation of differences between large numbers of $i_{l0}$ and $i_{l1}$ makes the impact of leakage on read access more severe. Figure 2 displays two extreme cases that consider the data pattern dependency of leakage current with 10K Monte Carlo (MC) sweeps. The orange lines represent V\textsubscript{RBL} of reading a `1' cell and all idle cells in the column to be `0' (maximum $i_{l}$), while the blue lines indicate V\textsubscript{RBL} of reading a `0' cell and the remaining cells being `1' (minimum $i_{l}$). It is challenging to distinguish voltage swings from different data cells with the existence of process variations. Therefore, additional auxiliary circuits are necessary to address the leakage noise caused by data differences. 

\subsection{Recent Works on Ultra-Low-Voltage SRAMs}
There have been several proposed techniques to enhance read access stability at low voltages. Do et al. \cite{do2012sensing} introduced a 9T SRAM cell that equalizes leakage current, avoiding the worst-case scenario. Chien et al. \cite{chien20180} proposed a 10T bitcell that enables robust operations at subthreshold voltages without the need for boost circuitry. Verma et al. \cite{verma2008256} employed sense amplifier redundancy to enhance sensing stability. However, for the widely used single-ended sensing scheme based on a domino-style circuit, these techniques may not be applicable. To address this, the author \cite{do201532kb} proposed a column-based randomization engine that randomizes data stored in the SRAM, resulting in the distribution of `1' and `0' in each column becoming close to 50\%, thus equalizing the leakage current on each column. However, this approach does not effectively solve the influence caused by the local mismatch between columns. Another proposed technique, as presented in \cite{wen2016bit}, is a PVT-tracking bias generator that compensates for the read bitline leakage to improve the sensing margin. This generator only provides compensation with a fixed value and increases the total energy consumption of the SRAM. Do et al. \cite{do20160} proposed a 9T cell with a leakage current calibration scheme, in which the leakage current on the RBL is sampled in the calibration stage and compensated in the reading stage. Meanwhile, in \cite{do2019energy}, a self-reference is set for each column so that the sense margin of the array is not affected by the local mismatch between columns. However, self-reference modules increase energy consumption. Finally, in \cite{shen2020modeling, ling2021design}, the error rate of the read operation is used as an indicator to adjust the SRAM working frequency. Unfortunately, this design is not suitable for ultra-low voltages where the bit error rate will drastically increase, making the cost of error correction intolerable.

\section{Analysis of Leakage and Sensing Window}
This section presents a model that describes the relationship between discharging delay, leakage current, read current, and supply voltages. This model aims to provide a better understanding of the principle behind Ultra8T SRAM.
\subsection{Modeling Sensing Window without Variations}
The parasitic capacitance on the RBL is denoted by $C_{RBL}$. Prior to a read operation, the RBL is pre-charged to V\textsubscript{DD}. Assuming that the accessed cell stores a `0’, the total discharge current on the RBL is the sum of the reading cell’s current $i_r$ and the leakage current $i_l$. Thus, the discharge delay, denoted by $T_{r0}$, of the RBL swinging from V\textsubscript{DD} to V\textsubscript{REF} can be expressed by \eqref{eq1},
\begin{equation}
    T_{r0}=C_{RBL}\frac{V_{DD}-V_{REF}}{i_l+i_{r0}} \label{eq1}
\end{equation}
As read current of reading a `1’ cell is cutoff current and can be ignored compared to leakage, discharge delay $T_{r1}$ can be defined similarly,
\begin{equation}
    T_{r1}\approx T_{leak}=C_{RBL}\frac{V_{DD}-V_{REF}}{i_l}\label{eq2}
\end{equation}
With the column depth of $N$, the leakage current $i_l$ is
\begin{equation}
    i_l=\sum_{k=1}^{K}i_{l0}^k+\sum_{q=1}^{M}i_{l1}^q\label{eq3}
\end{equation}
where $i_{l0}$ and $i_{l1}$ are cutoff current of the access transistor M7 in the unselected `0’ and `1’ cells, respectively; $K$ and $M$ represent the numbers of idle `0’ and `1' cells ($K+M=N-1$). From \eqref{eq3}, the leakage current is related to the ratio of `0’s and `1’s in the column (R\textsubscript{01}).

\begin{figure}
    \centering
    \subfigure[]{\includegraphics[width=0.8\linewidth]{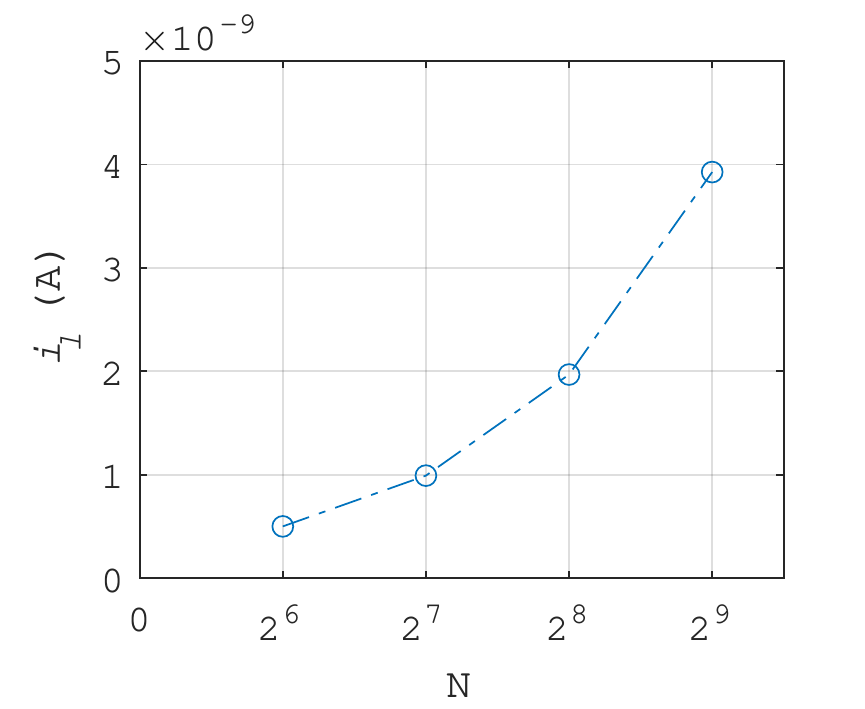} \label{fig_leak_N}}
    \subfigure[]{\includegraphics[width=0.8\linewidth]{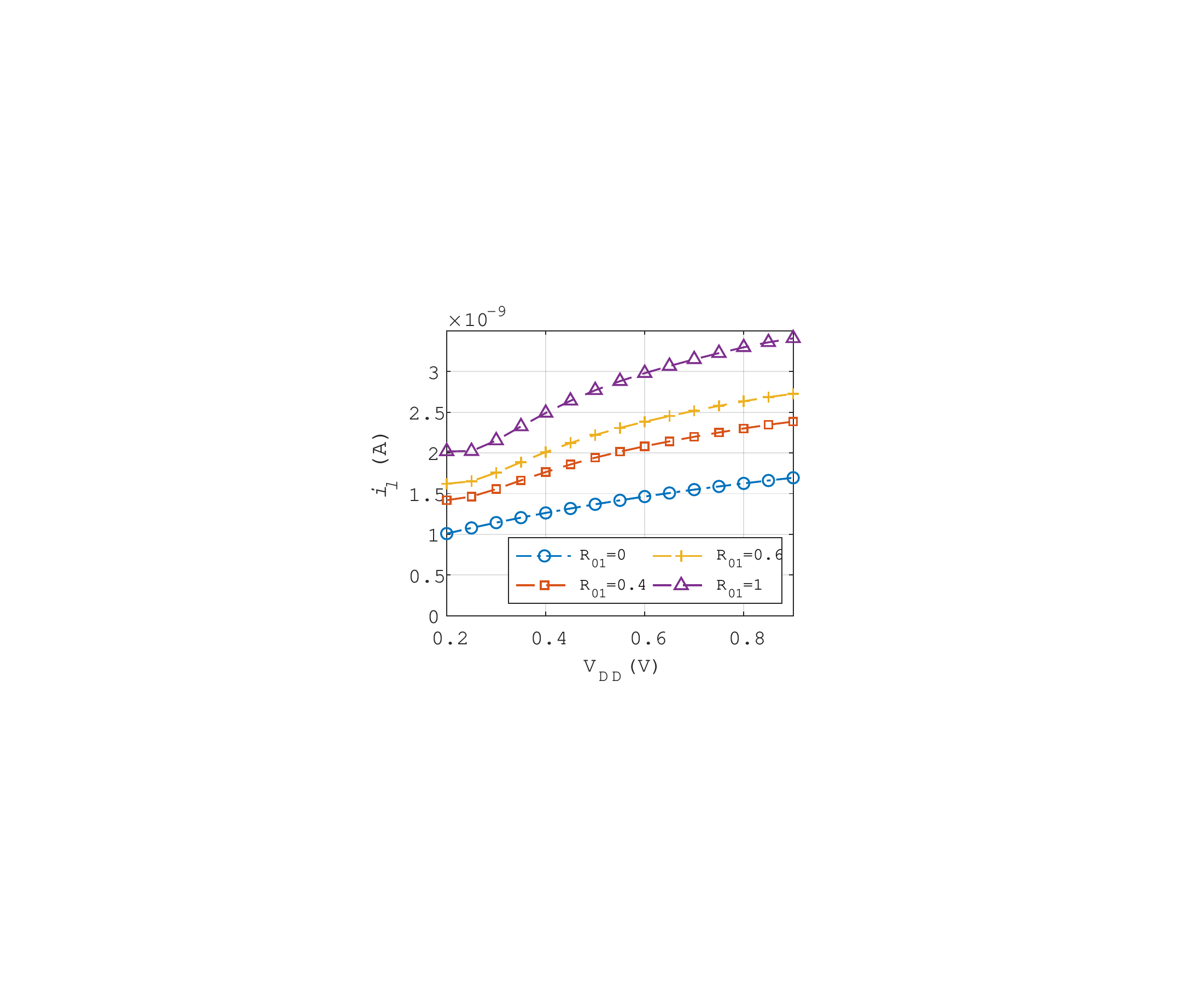} \label{fig_leak_VDD}}
    \subfigure[]{\includegraphics[width=0.8\linewidth]{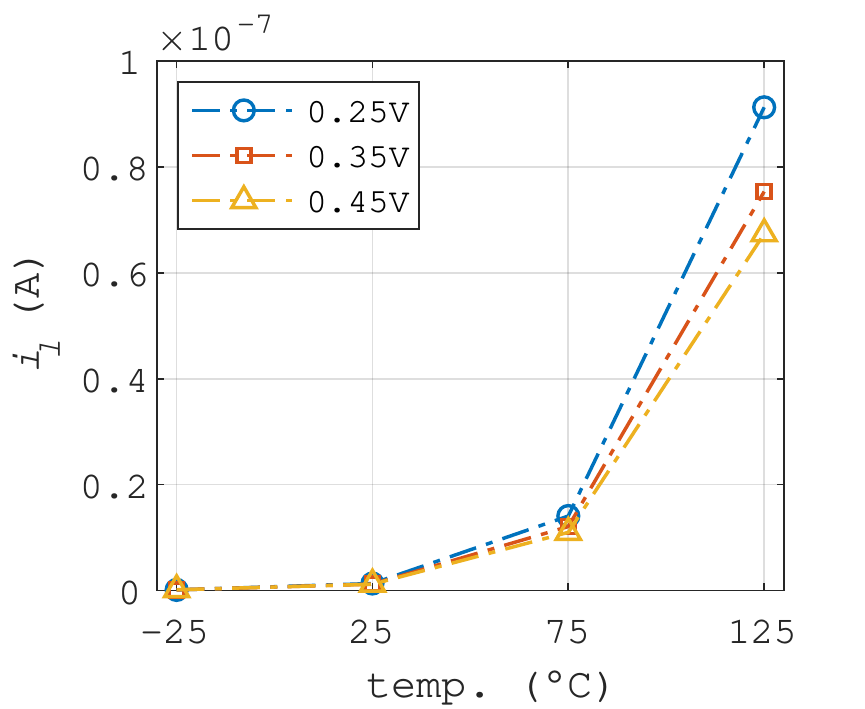} \label{fig_leak_temp}}
    \caption{(a)Mean of $i_l$ v.s. column-depth $N$ of SRAM at 0.25V, 25C. (b) Mean of $i_l$ v.s. V\textsubscript{DD} with different $R_{01}$. (c) $i_l$ v.s. temperature with different V\textsubscript{DD}.}
    \label{fig_leak}
\end{figure}

Here we define the safety sensing window in which the RBL discharging time $T_{RBL}$ during read access should be:
\begin{equation}
    \max(T_{r0})<T_{RBL}<\min(T_{leak}) \label{eq4}
\end{equation}
To quantify the $T_{RBL}$, the lower and the upper bounds in \eqref{eq4} should be specified explicitly. Equation \eqref{eqids} shows the classic transistor drain current model:
\begin{equation}
    I_{ds}=I_0e^\frac{V_{gs}-V_{th}}{nv_t}e^\frac{{\lambda V}_{ds}}{nv_t}(1-e^{-\frac{V_{ds}}{v_t}})\label{eqids}
\end{equation}
Here we assume there are no process variations in all SRAM cells for simplicity, in which equivalently treat $V_{th}$ in \eqref{eqids} as a fixed value. In a `0’ idle cell, M6 is turned on by the internally stored data and M7 stays off with a larger resistance. So, we can assume the voltage at the source of M7 is close to GND. Therefore, $i_{l0}$ can be expressed as
\begin{equation}
    i_{l0}=I_0e^\frac{-V_{th}}{nv_t} e^\frac{\lambda V_{RBL}}{nv_t}\left(1-e^{-\frac{V_{RBL}}{v_t}}\right)
    \label{eq6}
\end{equation}
In a `1’ idle cell, the voltage at the source of M7 is close to $V_{RBL}/2$ after assuming that M6 and M7 have the same on-resistance. Current $i_{l1}$ equals to
\begin{equation}
    i_{l1}=I_0e^\frac{-\frac{1}{2}V_{RBL}-V_{th}}{nv_t}e^\frac{\lambda V_{RBL}}{2nv_t}\left(1-e^{-\frac{V_{RBL}}{2v_t}}\right) \label{eq7}
\end{equation}
The drain current model can be simplified without introducing any transregional term due to its small value in the target V\textsubscript{DD} range. Combining \eqref{eq6} and \eqref{eq7} we can get
\begin{equation}
    i_{l1}=\alpha i_{l0}, \\
    \alpha=\exp\left(-\frac{\left(1+\lambda\right)V_{RBL}}{2nv_t}\right)<1
    \label{eq8}
\end{equation}
Then, according to \eqref{eq3} and \eqref{eq8}, leakage current equals to
\begin{equation}
    i_l=\left(K+\alpha M\right)i_{l0} \label{eq9}
\end{equation}
During read access, $i_l$ is a function of V\textsubscript{DD}, temperature, column-depth $N$, and $R_{01}$, as shown in Fig. \ref{fig_leak}. Leakage current $i_l$ has a linear relation with the column depth $N$. The decrease of $i_l$ slows down as V\textsubscript{DD} enters the sub-threshold regime, which is a non-linear function of V\textsubscript{DD}. And more `0’ cells in a column (higher $R_{01}$) account for larger $i_l$. Fig. \ref{fig_leak_temp} illustrates that $i_l$ increases exponentially with the temperature rise. All variable changes are consistent with the model \eqref{eq9}. 
In addition, the delay of RBL swing from V\textsubscript{DD} to V\textsubscript{REF} caused by leakage current is
\begin{equation}
    T_{leak}=C_{RBL}\frac{V_{DD}-V_{REF}}{\left(K+\alpha M\right)i_{l0}} \label{eq10}
\end{equation}

\begin{figure}
    \centering
    \subfigure[]{\includegraphics[width=0.8\linewidth]{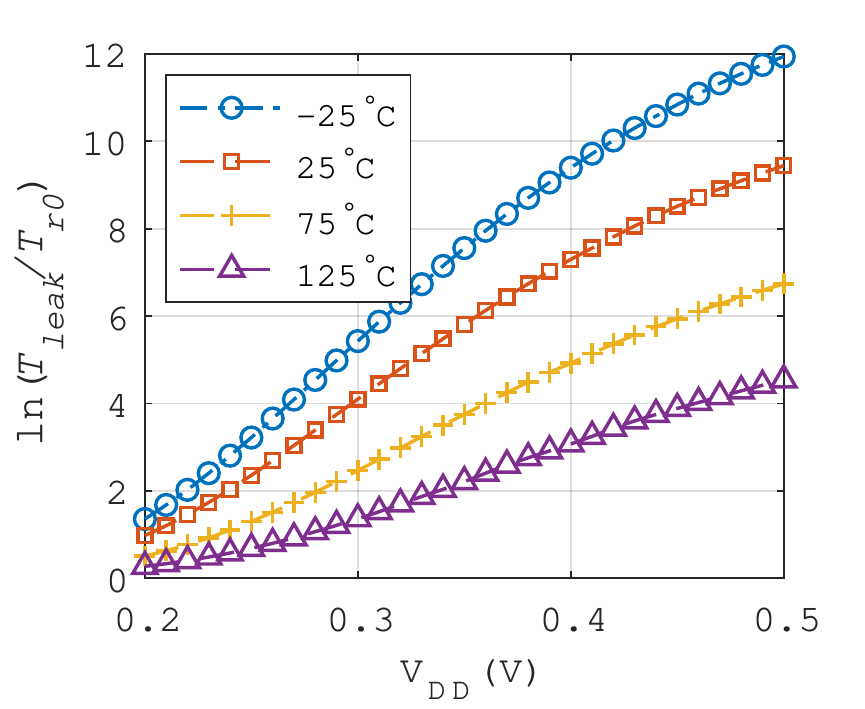} \label{fig_trate_vdd}}
    \subfigure[]{\includegraphics[width=0.8\linewidth]{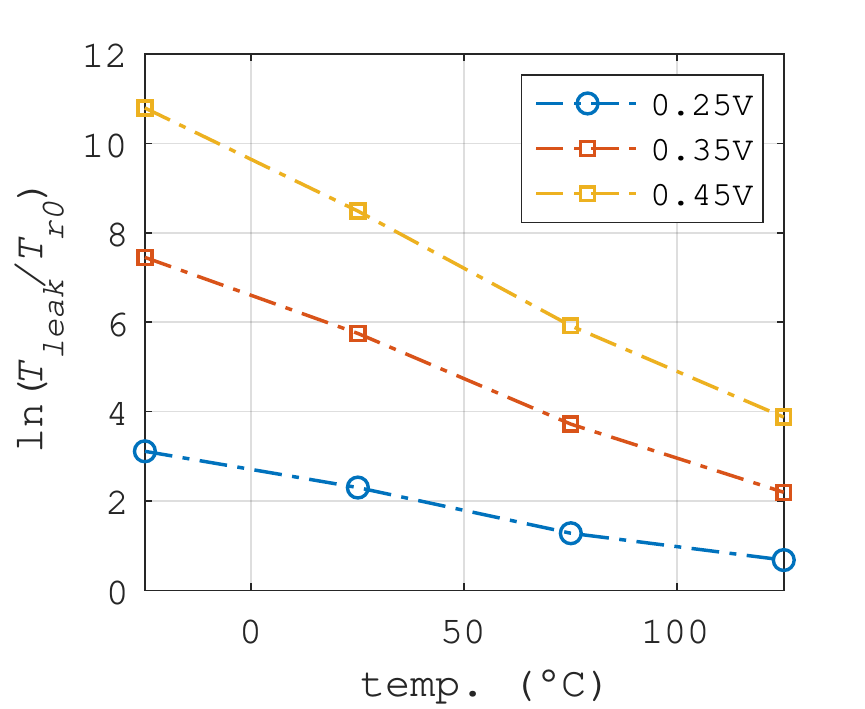} \label{fig_trate_temp}}
    \caption{(a) $\frac{T_{leak}}{T_{r0}}$ v.s. V\textsubscript{DD} with different temperatures. (b) $\frac{T_{leak}}{T_{r0}}$ v.s. temperature with different voltages.}
    \label{fig_trate}
\end{figure}

For reading a `0’ cell, the on-current of M7 can be expressed by 
\begin{equation}
    i_{r0}=I_0e^\frac{V_{DD}-V_{th}}{nv_t} e^\frac{\lambda V_{RBL}}{nv_t}\left(1-e^{-\frac{V_{RBL}}{v_t}}\right) \label{eq11}
\end{equation}
which is propotional to $i_{l0}$ in \eqref{eq6}. The discharge delay $T_{r0}$ can be re-written as
\begin{equation}
    T_{r0}=C_{RBL}\frac{V_{DD}-V_{REF}}{\left(e^\frac{V_{DD}}{nv_t}+K+\alpha M\right)i_{l0}}\label{eq12}
\end{equation}
According to \eqref{eq10} and \eqref{eq12}, $\max(T_{r0})$ can be achieved by minimizing the number of idle `0’ cells, $K=0$, in the meanwhile, $\min(T_{leak})$ can be achieved with the maximum `0’ idle cells, $K=N$. The relationship between the lower and upper bounds is
\begin{equation}
    \frac{\min(T_{leak})}{\max(T_{r0})}=\frac{1}{N}e^\frac{V_{DD}}{nv_t}+\alpha \label{eq13}
\end{equation}

Fig. \ref{fig_trate_vdd} clearly demonstrates that $T_{leak}/T_{r0}$ and V\textsubscript{DD} have exponential relationships at near-/sub-threshold voltages. Fig. \ref{fig_trate_temp} shows that $T_{leak}/T_{r0}$ also has a negative exponential relationship with the temperature at subthreshold voltages. Without process variation, we can define the safety sensing time as
\begin{equation}
    T_{RBL}=\beta\min\left(T_{leak}\right) \label{eq14}
\end{equation}
where the scale factor $\beta$ is a function of $N$, V\textsubscript{DD}, temperature. It should satisfy the following requirement,
\begin{equation}
    \left(\frac{1}{N}e^\frac{V_{DD}}{nv_t}+\alpha\right)^{-1}<\beta<1 \label{eq15}
\end{equation}
The min($T_{leak}$) can be measured by the existing SA circuits across the entire array so that the $T_{RBL}$ can be obtained after setting a proper value for $\beta$ (shown in Table \ref{tab1}).

\subsection{Combining with PVT Variations}
\begin{figure}
    \centering
    \includegraphics[width=0.8\linewidth]{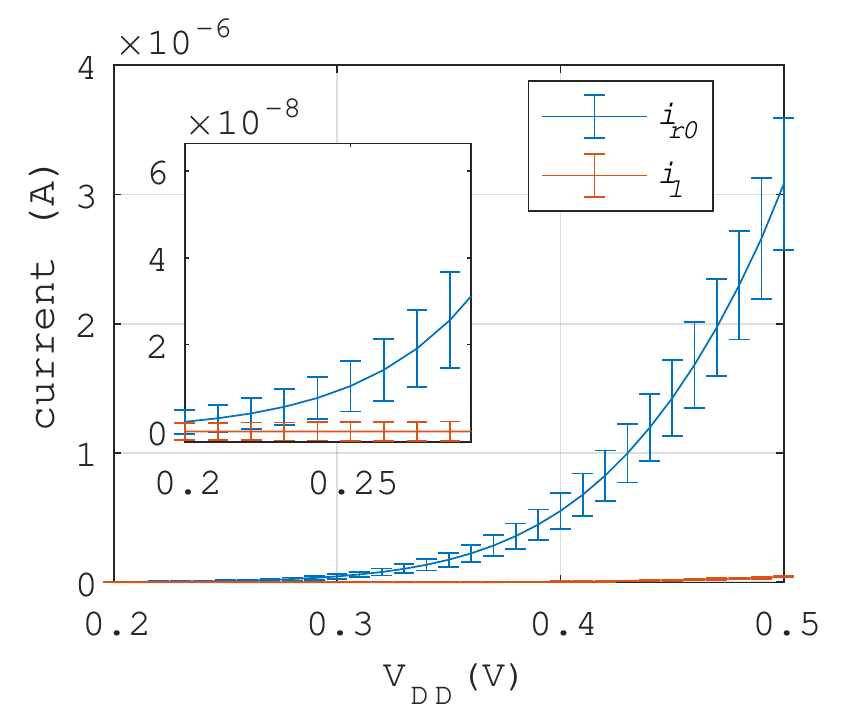}
    \caption{Mean and std of $i_{r0}$ and $i_l$ v.s. V\textsubscript{DD}.}
    \label{fig_ir_il_std}
\end{figure}

Now we take process variation into consideration. Due to the different magnitudes of on-and-off drain current of transistors, the influence of process variations differs. Fig. \ref{fig_ir_il_std} shows the mean and the triple of standard variation (std) of read and leakage current. It’s evident that read current experiences a larger variation than leakage current, and this difference grows exponentially as V\textsubscript{DD} scales up. To simplify the model, we still consider the idle cells as process-variation-free. The process variation is modeled as the threshold voltage variation. The variable $V_{th}$ in \eqref{eq11} is a normal random variable defined by the process technology, at the same time, $V_{th}$ in \eqref{eq6} can be treated as a fixed mean value of $V_{th}$ from all unaccessed cells. This causes only a minor change in $T_{r0}$ in \eqref{eq12}, leading to a reduced ratio of the lower and upper bounds in \eqref{eq13} 
\begin{equation}
    \frac{\min(T_{leak})}{\max(T_{r0})}=\frac{1}{N}e^{\frac{V_{DD}-\Delta V_{th}}{nv_t}}+\alpha \label{eq16}
\end{equation}
and $\beta$ in \eqref{eq15} changes into
\begin{equation}
    \left(\frac{1}{N}e^\frac{V_{DD}-\Delta V_{th}}{nv_t}+\alpha\right)^{-1}<\beta<1 \label{eq17}
\end{equation}
where $\Delta V_{th}$ is the absolute maximum difference of $V_{th}$ between the accessed cell and the unaccessed cell in the array. These changes in \eqref{eq16} and \eqref{eq17} demonstrate that the existence of process variations makes the boundary of $\beta$ stringent and shrinks the safety sensing window. 

At last, the total read delay $D$ equals
\begin{equation}
    D=T_{RBL}+T_{SA}+T_{RWL} \label{eqd}
\end{equation}
where $T_{SA}$ is the delay of SA activation, and $T_{RWL}$ is the delay of raising the RWL to V\textsubscript{DD}. According to \eqref{eq14}, $T_{RBL}$ and working frequency can be set based on the minimum of $T_{leak}$ and a proper $\beta$. We will implement the ideas of the above model in the next section.

\section{Design of Ultra8T SRAM}
This section introduces the principle of Ultra8T SRAM and its detailed implementation to enhance ultra-low-voltage read operations. 

\subsection{The Principle}\label{AA}

\begin{figure}[tb]
\centering
\subfigure[]{\includegraphics[width=0.8\linewidth]{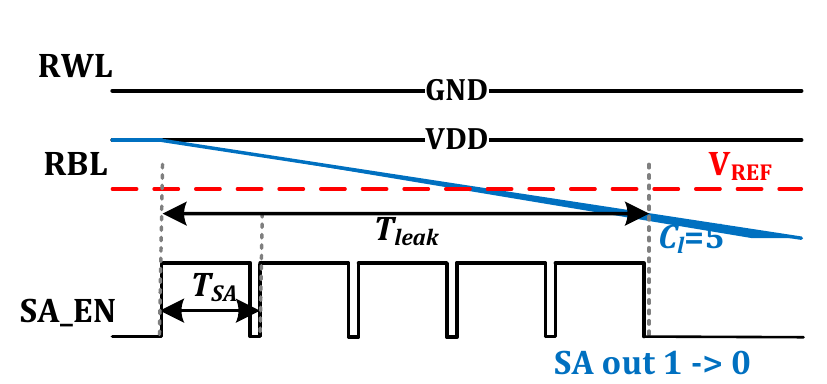} \label{fig3-a}}
\subfigure[]{\includegraphics[width=0.8\linewidth]{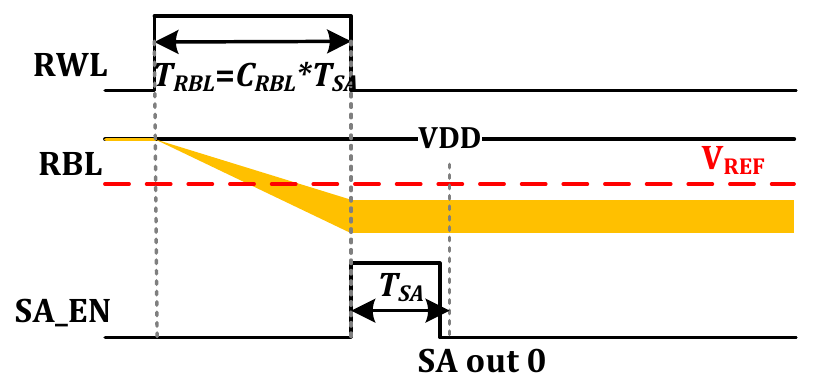} \label{fig3-b}}
\caption{Timing diagram of Ultra8T in (a) test mode and (b) read mode.}\label{fig3}
\vspace{-0.4cm}
\end{figure}

The basic concept is to establish the function relationship between the leakage current and the possible delay of RBL discharging (described by \eqref{eq14}). Leakage current, which is an analog variable affected by V\textsubscript{DD}, temperature, column-depth $N$, and $R_{01}$ in a column, can be converted to digital values ($T_{leak}$) by the existing SA circuits. By detecting the leakage current in advance, the upper bound of $T_{RBL}$ can be quantified (i.e., the safety sensing window at the current PVT is determined under the current SRAM configuration). The relation in \eqref{eq14} can be implemented as a lookup table to minimize the hardware overhead.
To establish the table, Ultra8T enters a test mode before normal access. As shown in Fig. \ref{fig3-a}, RBL is first pre-charged to V\textsubscript{DD} and starts discharging due to the existence of leakage current. SAs are activated by the SA\_EN signal several times, while all RWLs remain low. The sensing process will not stop until the data words read by SAs have changed. Leakage counters in the SRAM controller record the number of sensing cycles 
\begin{equation}
    \min(T_{leak}) = c_L\bullet T_{SA}
\end{equation}
to represent the value of $\min(T_{leak})$. As the proposed SRAM is organized as a bank, we only record the worst-case after scanning all columns in the bank.
The RBL discharging time of Ultra8T SRAM bank is defined as 
\begin{equation}
    T_{RBL} = c_{R}\bullet T_{SA} < \min(T_{leak})
    \label{eq18}
\end{equation}
where $c_{R}$ is stored in the lookup table as a value and queried using a key, $c_L$. As $T_{leak}$ is affected by PVT, we traverse all possible V\textsubscript{DD} to find V\textsubscript{DDMIN} and record $c_L$ and $c_{R}$. 

\begin{table}[!bp]
\begin{center}
\caption{The lookup table obtained by simulation at 25$^\circ$C}\label{tab1}
\begin{tabular}{l|lll}
\hline
V\textsubscript{DD}(V) & $c_L$ & $c_{R}$ & $\beta$ in \eqref{eq14} \\ \hline
0.2    & 4        & Invalid &  1.1 (Invalid) \\
0.25   & 13       & 6       &  0.45  \\
0.3    & 38       & 6       &  0.14  \\
0.35   & 109      & 5       &  0.044 \\
0.4    & 270      & 5       &  0.016 \\
$\leq$ 0.45   & 512 & 4     &  0.0073 \\ \hline
\end{tabular}
\end{center}
\end{table}

Table \ref{tab1} shows an example of the lookup table obtained by simulation at 25$^\circ$C. The $c_{R}$ reaches its minimum value at 0.45V, thus, there is no need to track $c_{L}$. The values of the lookup table can be pre-written through either simulation or offline testing (like Build-in Self-Test, BIST).
%During the generation of Table 1, $T_{SA}$ is configured using a PVT-tracking internal clock, as explained in Section \ref{s3-time}. 
%After the voltage change, min($T_{leak}$) is measured by means of pulse counting to obtain $c_{L}$' at the current voltage. The two high levels of $c_{L}$' are compared with the two high levels of $c_{L}$ in the lookup table. If they are equal, the corresponding $C_{RBL}$ is read out and the SRAM timing is configured.

\subsection{Overall Structure}

\begin{figure}[tb]
\centering
\includegraphics[width=1\linewidth]{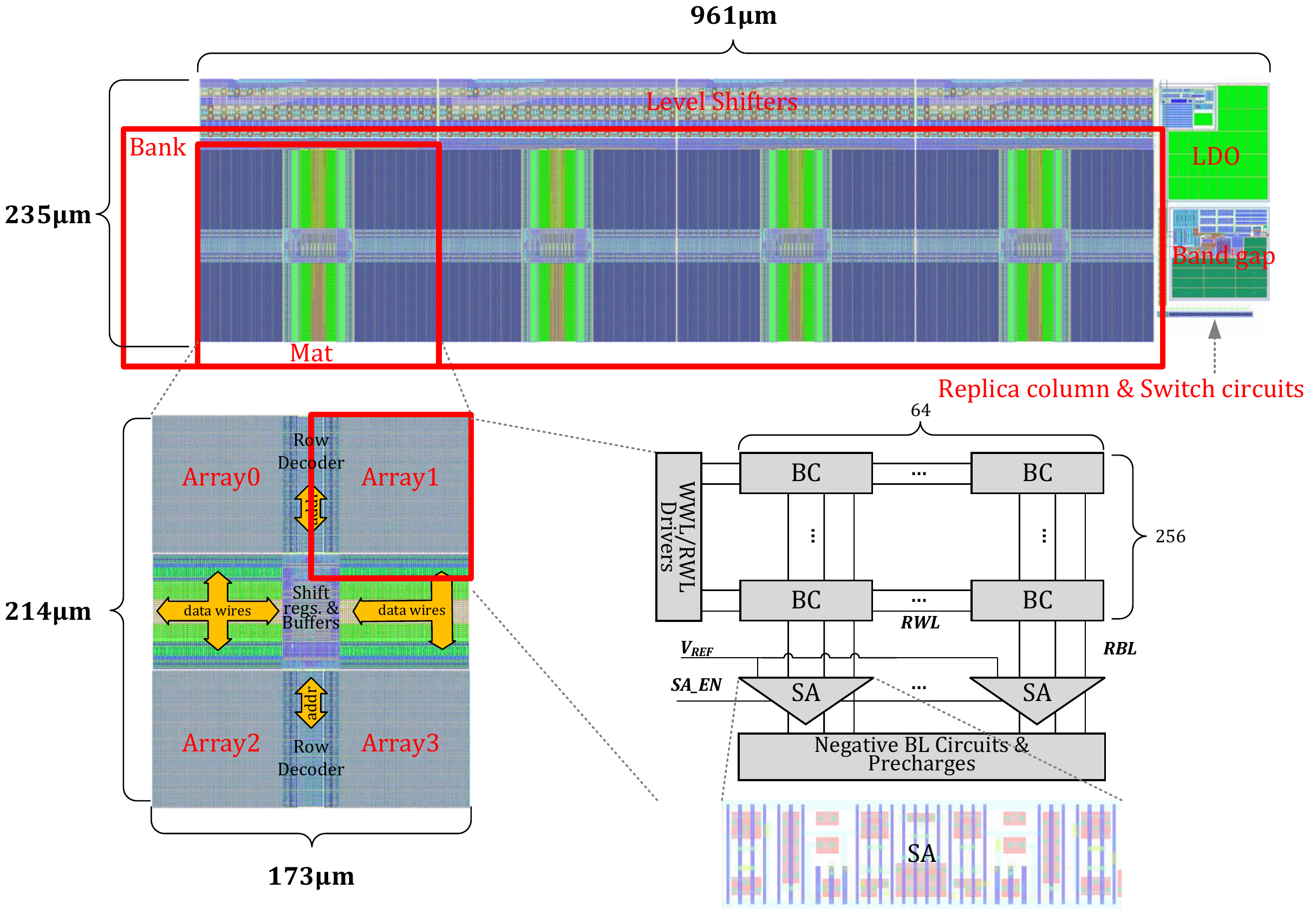} 
\caption{Overall structure of Ultra8T SRAM.}\label{fig4}
\end{figure}

Figure \ref{fig4} depicts that the Ultra8T macro includes an SRAM bank, a digitized timing module (comprised of replica columns, switch circuits, and shift registers), and the analog part. 
The digitized timing generates a basic internal clock for Ultra8T SRAM. All widths of signal pulses are multiples of the clock period (e.g. pulses in \eqref{eq18}). 
The LDO and bandgap are configurable, with the default V\textsubscript{REF} = V\textsubscript{DD}/2. 
Ultra8T is organized as a mat structure, which includes 4 SRAM arrays with a size of 256$\times$64. Each mat has shift registers to configure the signal pulses to different lengths, and only one array is activated during access. Two arrays share a row decoder, and 4 data arrays share a 64-bit width data port.
The lookup table is integrated into the SRAM controller, which is not shown in Fig. \ref{fig4} for simplicity. It is implemented as a register file to enable sub-threshold operations. A 10-bit saturated counter is used to track $c_{L}$ at different PVT. 

An SRAM array consists of write/read word-line drivers, SAs, negative BL circuits, and precharges. The SAs in Ultra8T need to satisfy two requirements of the implementation due to the existence of the test mode. Firstly, the SA should have a large input impedance to prevent RBL discharge through it. Secondly, the offset voltage of the SA should be minimized to obtain accurate sensing cycles $c_{L}$ and avoid a large voltage swing on RBL to save energy. Ultra8T employs a single-ended offset-canceling SA (SOSA) based on a voltage-latch type SA \cite{sosa}, which will be discussed in Section \ref{s3-sosa}.

On the other hand, due to the leakage detection strategy where the voltage swing of RBL is converted into digital timing pulses, a PVT-tracking variation-suppressed internal clock is also required. Section \ref{s3-time} introduces the design of the digitized timing module.

\subsection{SOSA}\label{s3-sosa}

\begin{figure}[tb]
\centering
\includegraphics[width=1\linewidth]{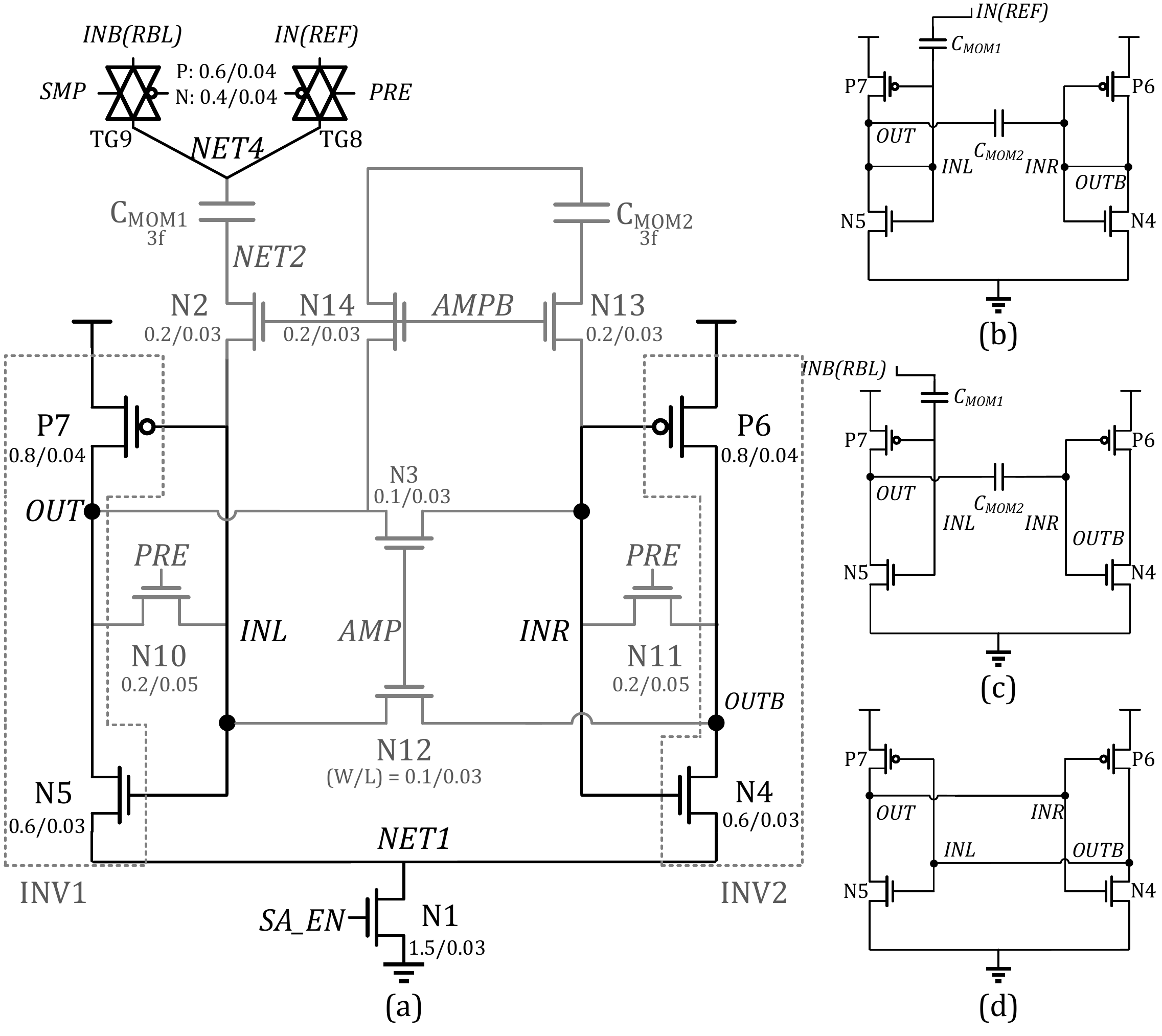} 
\caption{(a) Overall circuits of SOSA \cite{sosa}. (b) Precharge phase, (c) voltage sampling phase, and (d) amplifying phase in a read operation.}\label{fig5}
\vspace{-0.4cm}
\end{figure}

The sense amplifier requires a minimum differential input signal larger than the input-referred offset voltage  (V\textsubscript{OS}) to make a reliable decision. A larger standard deviation of offset voltage distributions ($\sigma_{OS}$) has a huge negative effect on read speed and yield. 
%The trip point (V\textsubscript{TRIP}) is defined as the input voltage that equals the output voltage of an inverter. The offset voltage is mainly caused by V\textsubscript{TRIP} mismatch between the cross-coupled inverters of an SA (e.g., INV1 and INV2 in Fig. 9) due to V\textsubscript{TH} variations. PVT variations under nanoscale technology would further exacerbate this mismatch. 

Ultra8T SRAM adopts a single-ended offset-canceling SA (SOSA) \cite{sosa} to reduce V\textsubscript{OS}. The overall circuitry of SOSA is shown in Fig. \ref{fig5}(a). SOSA is constructed from a basic voltage latch-type SA structure (INV1 and INV2). In addition, it is equipped with 2 capacitors, C\textsubscript{MOM1}, to couple the input voltage swing into the internal sensing nodes, and C\textsubscript{MOM2}, to store the V\textsubscript{TRIP} mismatch between the 2 inverters. Different from other symmetric SAs, the basic principle of SOSA is transmitting a small signal from the left to the right inverter sequentially. Thereby, it is very sensitive to the input voltage changes and has a very small V\textsubscript{OS} across a wide voltage range. There are 3 phases in operating SOSA, including precharge, voltage sampling, and amplifying. In the precharge phase (Fig. \ref{fig5}(b)), TG8 connects NET4 to a voltage reference, N2, N10, N11, N13, and N14 are switched on by PRE and AMPB. This makes V\textsubscript{INL} = V\textsubscript{TRIP1}, and V\textsubscript{INR} = V\textsubscript{TRIP2}. The voltage difference between IN and INL is stored by C\textsubscript{MOM1}, meanwhile, the difference between INL/OUT and INR is stored by C\textsubscript{MOM2}. In the voltage sampling phase (Fig. \ref{fig5}(c)), TG8 is off, TG9 is turned on by SMP, N2, N13, and N14 remain in their active states. The voltage difference between RBL and the reference is passed through C\textsubscript{MOM1} to INV1, then to INV2 through C\textsubscript{MOM2}. In the amplifying phase (Fig. \ref{fig5}(d)), only N3 and N12 are activated by AMP to configure SOSA to latch mode. The footer N1 is always turned on by SA\_EN during the entire operation.

\begin{figure}[tb]
\centering
\includegraphics[width=0.8\linewidth]{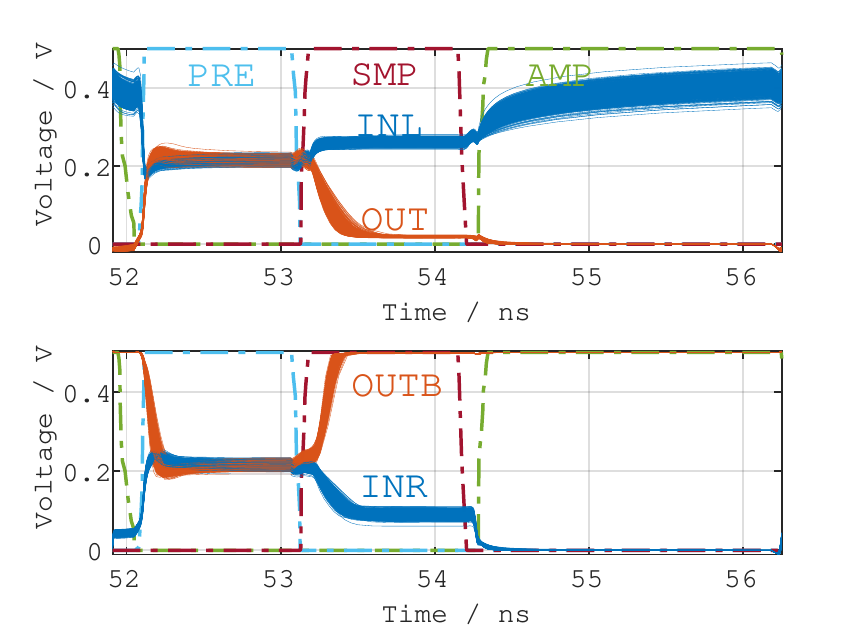} 
\caption{Timing diagram of SOSA \cite{sosa}, assuming V\textsubscript{IN} $<$ V\textsubscript{INB}.}\label{fig6}
%\vspace{-0.4cm}
\end{figure}

Due to the precharge operation, the 2 inverters are all in the metastable state with the largest gain factors so that any small stimulation at the node INL and INR can be amplified quickly. This feature is leveraged by the voltage sampling phase, providing a very small V\textsubscript{OS} across a wide voltage range. The timing diagram collected from 1K Monte Carlo (MC) sweeps is shown in Fig. \ref{fig6}. The lengths of precharge, voltage sampling, and amplifying phases are set to 2, 2, and 4 clock periods, respectively.

\subsection{Digitized Timing}\label{s3-time}
An inaccurate SA enabling (SA\_EN) signal can lead to inaccurate leakage detection of Ultra8T. In order to suppress timing variation and track the PVT condition, a digitized timing scheme is proposed. Similar to the design \cite{yang2018double}, it is comprised of 3 parts, switch circuits, two 256-depth replica 8T columns (same as the column depth of the array), and configurable shift registers that reside in the timing configuration circuitry in the mat structure. 

\begin{figure}[tb]
\centering
\includegraphics[width=1\linewidth]{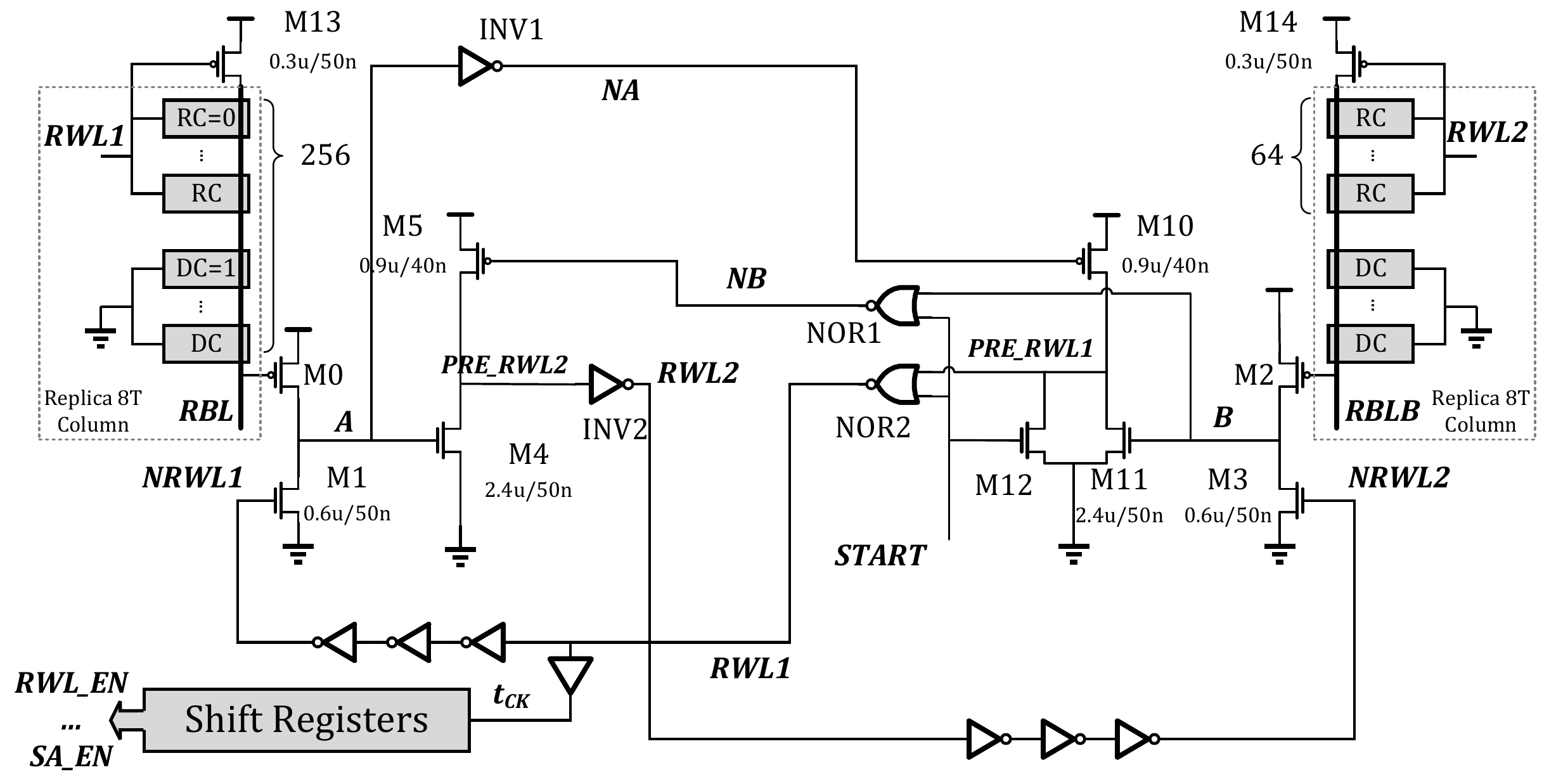} 
\caption{The switch circuits in digitized timing.}\label{fig7}
\end{figure}

\begin{figure}[tb]
\centering
\includegraphics[width=1\linewidth]{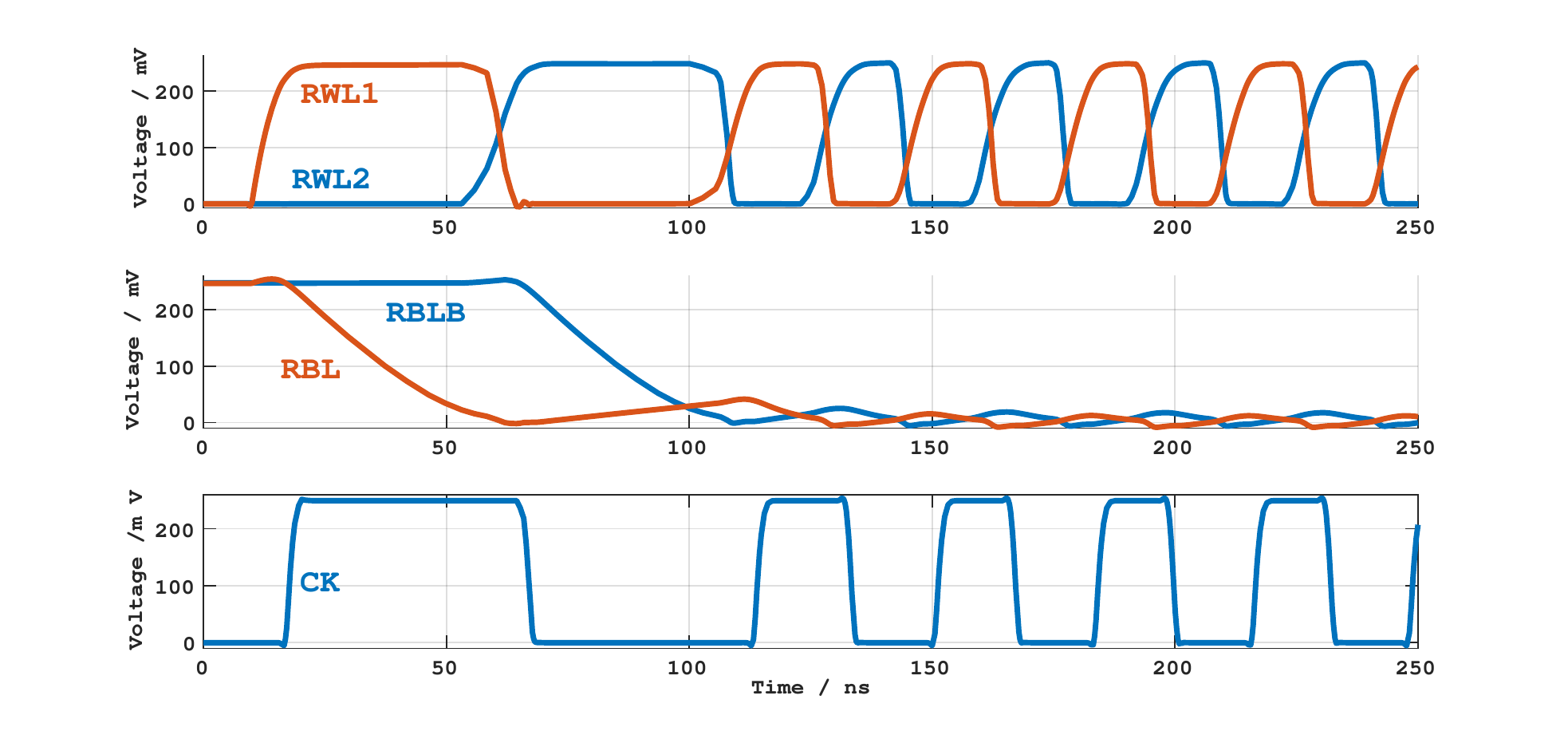} 
\caption{The switch circuits in digitized timing.}\label{fig8}
\end{figure}

As shown in Fig. \ref{fig7}, the RWLs of the replica cells (RCs) in the 2 columns connect to RWL1 and RWL2 of the switch respectively. Data of RCs are pre-set to `0' and the remaining dummy cells (DCs) are pre-set to `1'. The column depth is 256 and the number of RCs is 64. In the reset state, the START signal is high, V\textsubscript{RWL1} and V\textsubscript{NB} are `0's, making PMOS M5 active and V\textsubscript{RWL2} = `0'. Hence, PMOS M13 and M14 are activated, charging both RBL and RBLB to V\textsubscript{DD}. Transistors M10 and M11 are off, and node PRE\_RWL1 is pulled down to the ground by M12. When the START signal is `0', in the 1st step, RWL1 becomes high, and RBL starts discharging through the RCs. When RBL is discharged to the voltage that can turn on M0, node A is pulled up to V\textsubscript{DD}, resulting in an activated M4 and high V\textsubscript{RWL2}. At the same time, M10 is turned on to charge PRE\_RWL1, then RWL1 returns to low, ending the discharge of RBL and recharging RBL to high. In the 2nd step, with V\textsubscript{RWL2}=`1', RBLB starts discharging through the other RCs. Similar to the first step, when V\textsubscript{RBLB} is sufficient to turn M2 on, node B will be pulled up to V\textsubscript{DD}. Then V\textsubscript{PRE\_RWL1} = `0', RWL1 is asserted again. At the same time, M5 makes V\textsubscript{PRE\_RWL2} = `1', ending the discharge of RBLB. Then, the switch circuits go back to step 1 again. Fig. \ref{fig8} shows the diagram of the switch.

The circuit structure of the switch is symmetrical, and the internal signals are cyclically flipped in a periodic manner generating a clock signal. Since both RBL and RBLB are simultaneously discharged through 64 replica cells, the delay deviation of CK is greatly reduced. Additionally, shift registers comprised of multiple D-flipflops are used to generate different control signals with different pulse widths. They are implemented individually for each mat and integrated into the center of the mat (Fig. \ref{fig4}).

\section{Evaluation}
All the aforementioned designs have been validated in TSMC’s 28nm commercial technology. We have collected all performance metrics by conducting 1K Monte Carlo sweeps, with all local variations turned on, at a temperature of 25$^\circ$C. SSG, TTG, and FFG represent different types of global process corners, slow NMOS slow PMOS, typical NMOS typical PMOS, and fast NMOS fast PMOS respectively. 
The post-layout netlist has been generated using StarRC, and the simulations have been carried out using HSPICE. In order to speed up the simulations, we have separately simulated each component, including the timing module, a single SRAM array, and the decoder. All data presented in this section have been collected from the post-layout simulations. Note that the delay and energy overhead of the controller and analog circuits have not been included. To achieve the ideal V\textsubscript{DDMIN}, the test data has been set at $R_{01}$ = 50\%. More complex data patterns or write randomization techniques will be left for future work.

\subsection{Normal Read Mode}

\begin{figure*}[tb]
\centering
\subfigure[]{\includegraphics[width=0.3\linewidth]{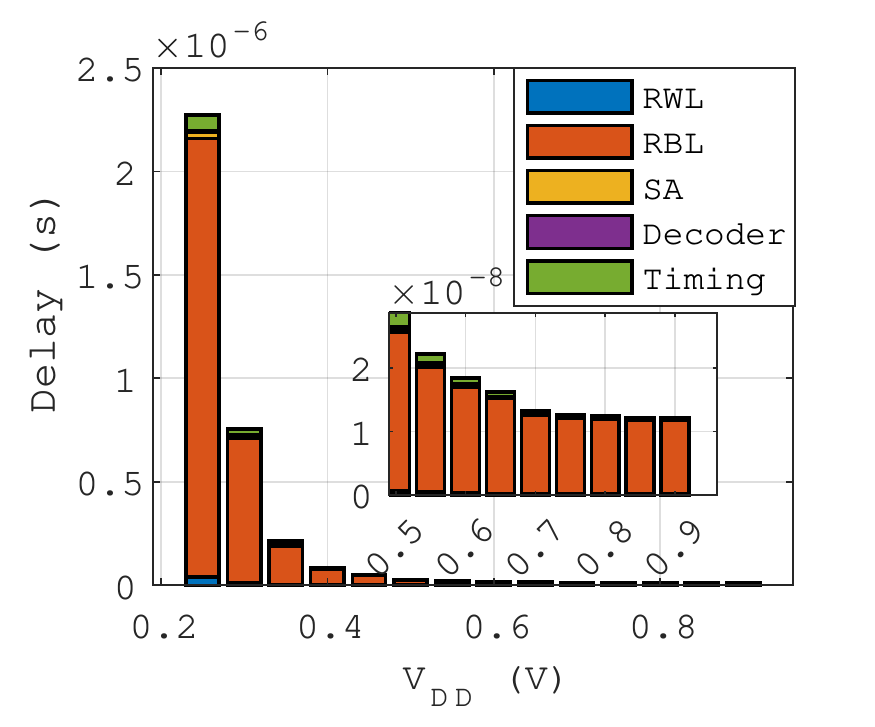} \label{fig9-a}}
\subfigure[]{\includegraphics[width=0.3\linewidth]{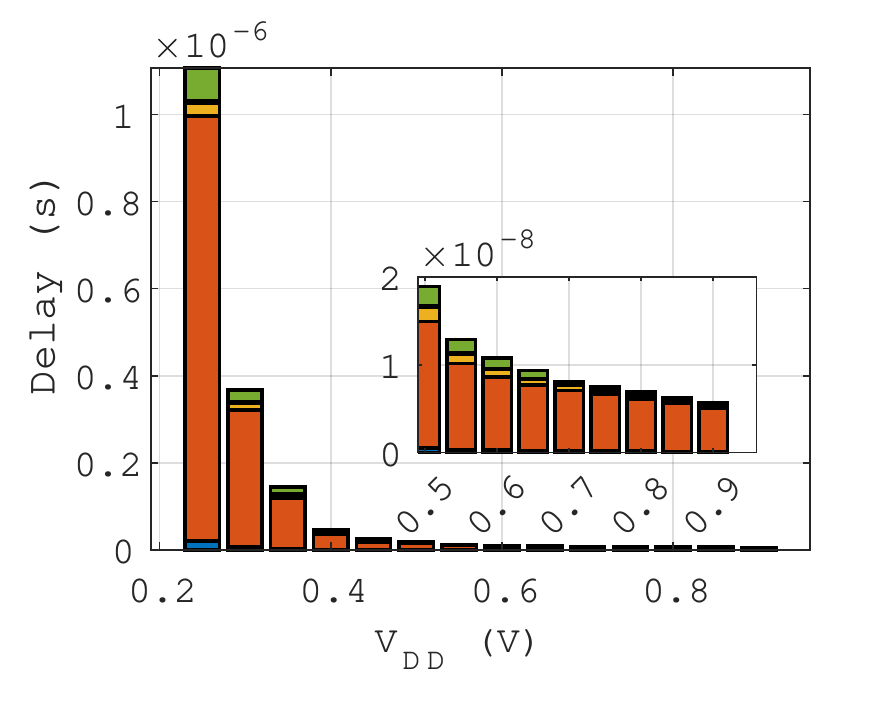} \label{fig9-b}}
\subfigure[]{\includegraphics[width=0.3\linewidth]{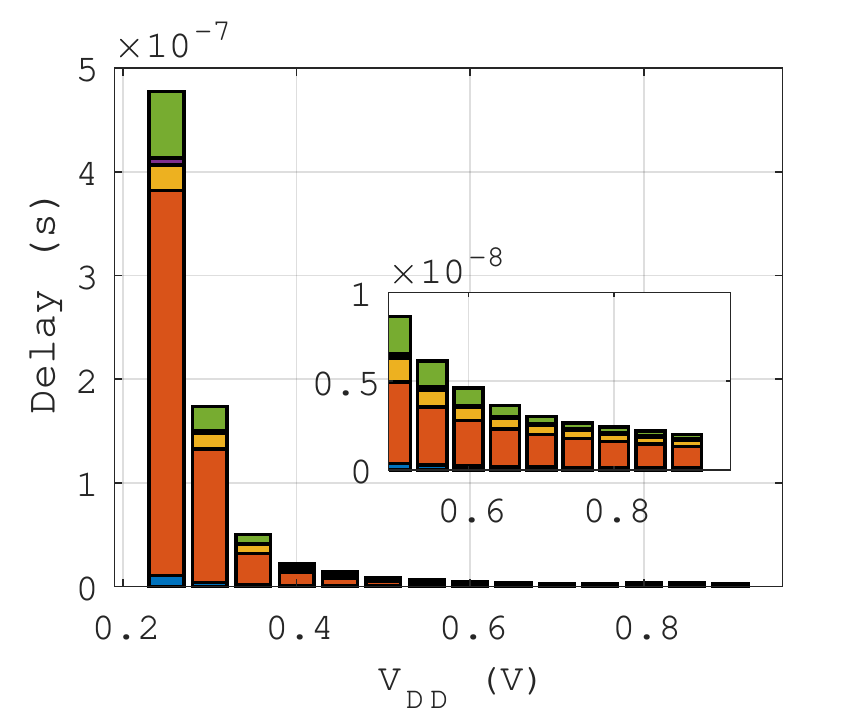} \label{fig9-c}}
\caption{Delay decompositions v.s. V\textsubscript{DD} at 25$^\circ$ C (a) SSG, (b) TTG, and (c) FFG.}\label{fig9}
%\vspace{-0.4cm}
\end{figure*}

\begin{figure*}[tb]
\centering
\subfigure[]{\includegraphics[width=0.3\linewidth]{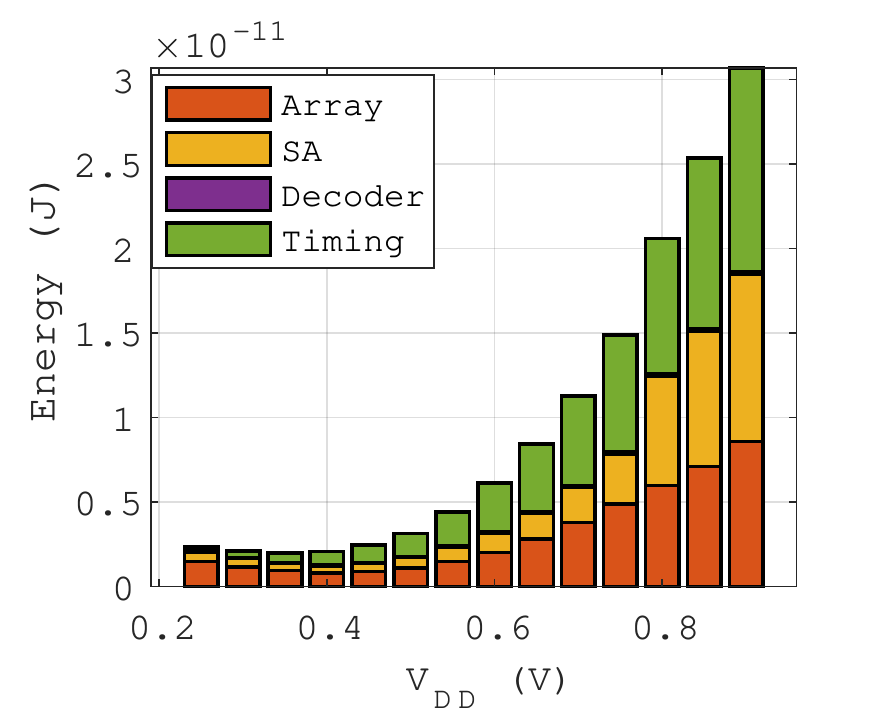} \label{fig10-a}}
\subfigure[]{\includegraphics[width=0.3\linewidth]{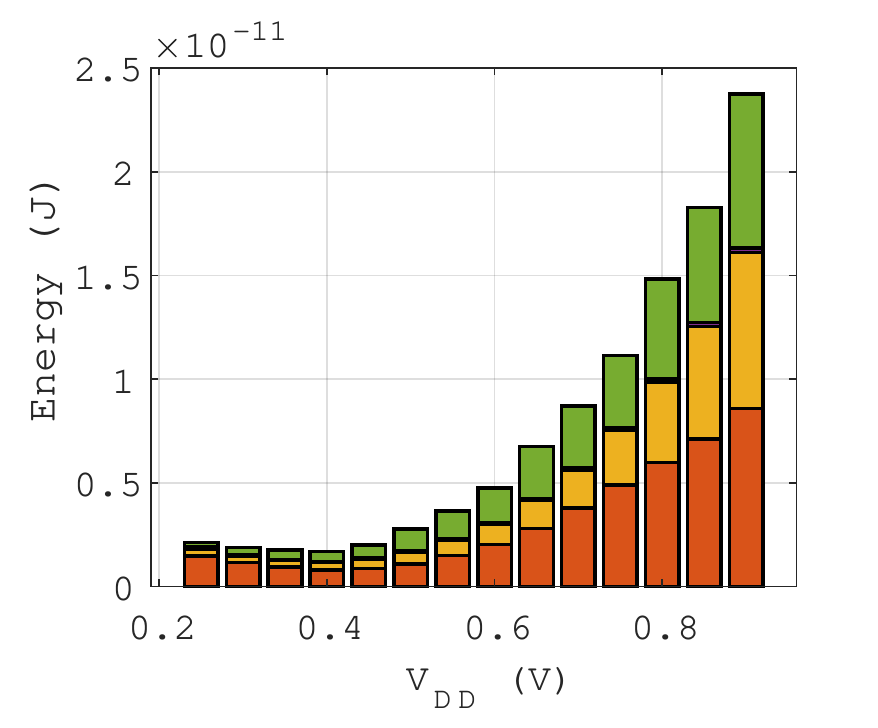} \label{fig10-b}}
\subfigure[]{\includegraphics[width=0.3\linewidth]{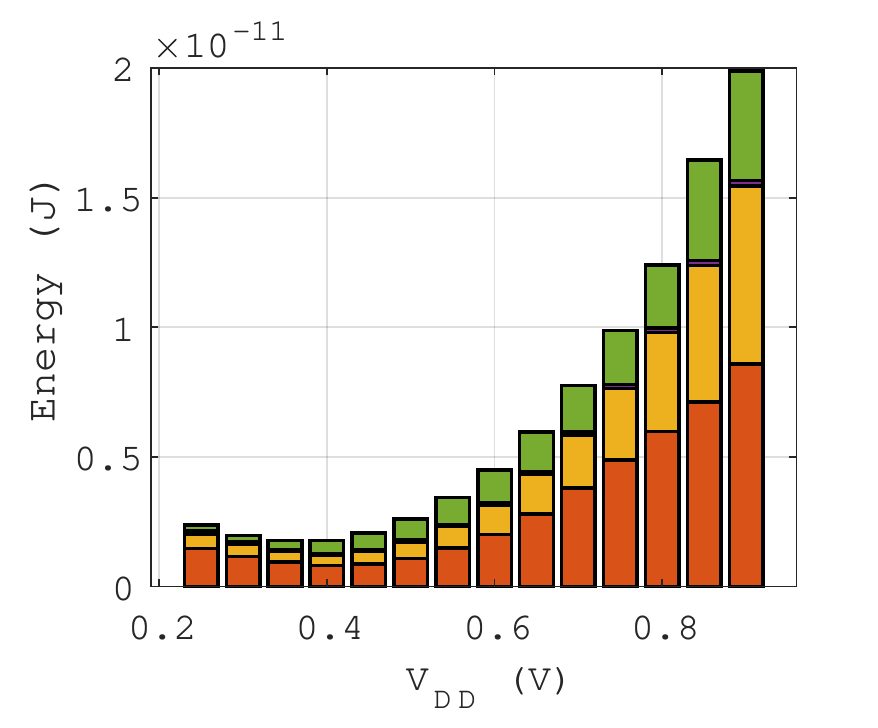} \label{fig10-c}}
\caption{Energy decompositions v.s. V\textsubscript{DD} at 25$^\circ$C (a) SSG, (b) TTG, and (c) FFG.}\label{fig10}
%\vspace{-0.4cm}
\end{figure*}

Fig. \ref{fig9} depicts the read delay decomposition. As V\textsubscript{DD} scales down, the delay increases exponentially. The read delay of Ultra8T is 1.1\textmu s at 0.25V and 24.8ns at 0.45V. At low V\textsubscript{DD}, the SRAM array dominates the overall access delay. Long $T_{RBL}$ significantly slows down the SRAM frequency at ultra-low voltages. When the process corner becomes better, the proportion of $T_{RBL}$ in the read delay is reduced. 
Figure \ref{fig11} compares the read energy over a wide range of V\textsubscript{DD}. The array consumes a significant portion of the overall energy at all operation voltages due to the leakage power. This figure also shows that the minimum energy point of the proposed SRAM is within the range of 0.25V to 0.45V. However, this point may vary in different applications that have different $R_{01}$ and SRAM configurations. 
We find the best energy-delay product (EDP) of Ultra8T is achieved at 0.45V.

\subsection{Test Mode}

\begin{figure}[tb]
\centering
\subfigure[]{\includegraphics[width=0.8\linewidth]{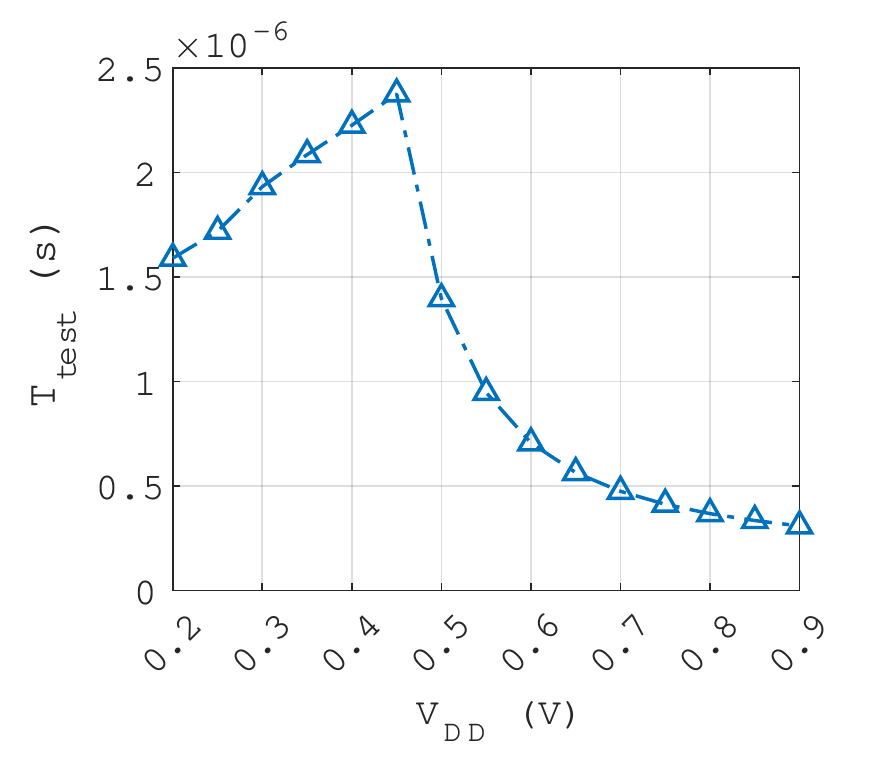} \label{fig11-a}}
\subfigure[]{\includegraphics[width=0.8\linewidth]{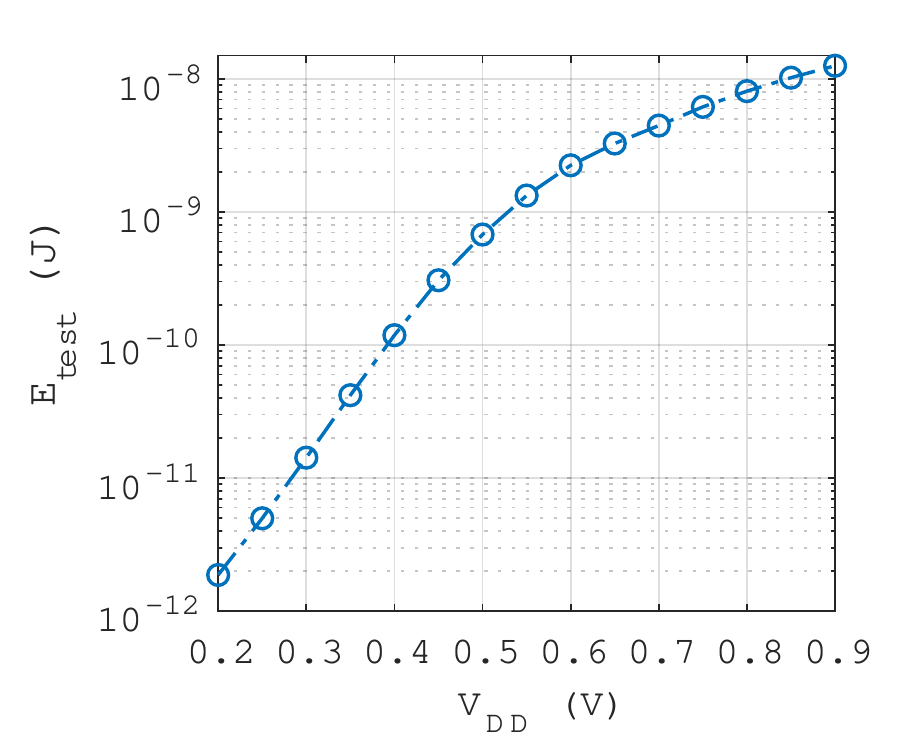} \label{fig11-b}}
\caption{(a) Delay and (b) energy of test mode v.s. V\textsubscript{DD}}\label{fig11}
%\vspace{-0.4cm}
\end{figure}

Figure \ref{fig11-a} shows the test delay (T\textsubscript{test}) presents a turning point. In the first phase, T\textsubscript{test} at subthreshold voltages increases linearly with the voltage. In the second phase, the T\textsubscript{test} starts reducing.
This is because the leakage counter gets saturated at 0.45V, in the meanwhile, the frequency of digitized timing increases as the supply voltage scales up.
Test energy is larger than that in normal read mode because the sense amplifiers (SAs) are continuously activated. Additionally, the test mode requires test-data writing. In Fig. \ref{fig11-b}, we show that the energy overhead of the test mode increases as the voltage increases. Since leakage counter $c_L$ is saturated in the nominal voltage range, the uptrend of energy overhead in the test mode slows down.
However, these operations can be executed only once for each SRAM module with the data write randomization technique, like \cite{chien20180}. In our future work, we will explore an energy-efficient run-time testing scheme. 

\subsection{SOSA}

%\begin{table}[tb!]\scriptsize
%\begin{center}
%\caption{Standard Deviation of Offset Voltage of SOSA}\label{tab2}
%\begin{tabular}{l|l|l|l} \hline
%VDD /V & SSG & TTG & FFG \\ \hline
%0.2 & 11.4 & 10.1 & 10.0 \\
%0.5 & 5.4 & 5.9 & 7.1 \\
%0.8 & 6.1 & 6.0 & 5.8 \\ \hline
%\end{tabular}
%\end{center}
%\end{table}

\begin{figure}[tb]
\centering
\includegraphics[width=0.8\linewidth]{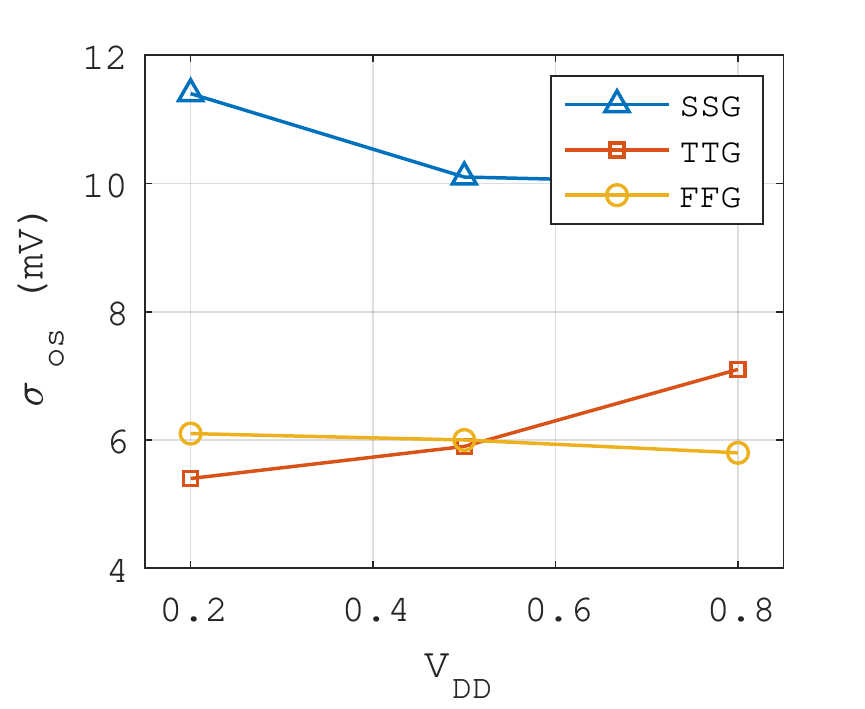} 
\caption{Standard deviation of offset voltage of SOSA.}\label{fig_vos}
\end{figure}

The layout depicted in Fig. \ref{fig4} is completely symmetrical about N1 and N14. The area overhead is 4.9 \textmu m$^2$. To save space while maintaining the SRAM column pitch, two MOM capacitors are stacked on the transistors. In Fig. \ref{fig_vos}, $\sigma_{OS}$ is consistently below 11.4 mV. We discovered that the precharge phase accounts for nearly half of the total energy consumed by SOSA in the simulation. However, this issue can be mitigated by employing fine-grained timing control or multiplexers.

\subsection{Digitized Timing}
\begin{figure}[tb]
\centering
\includegraphics[width=0.8\linewidth]{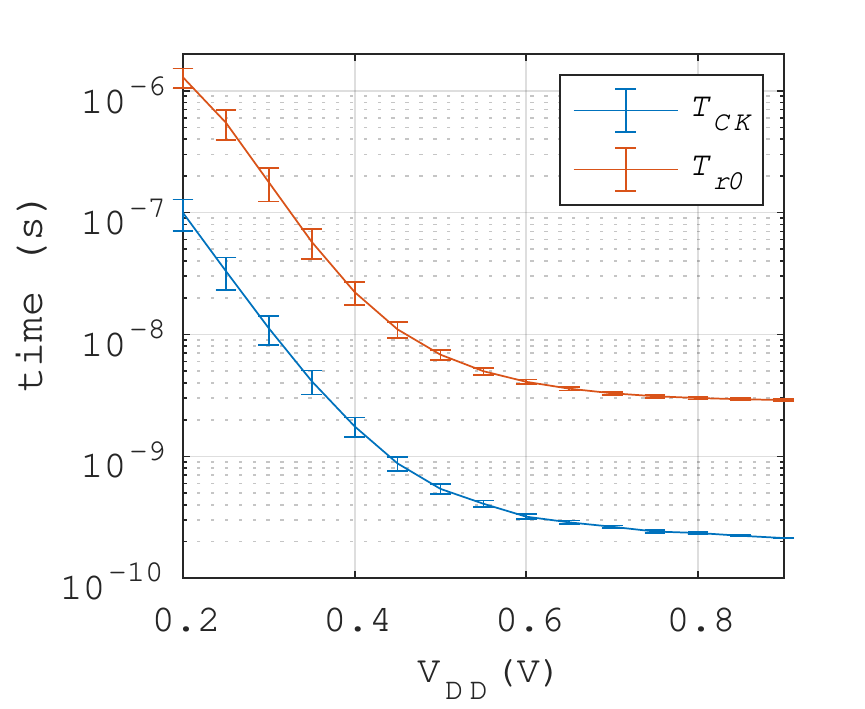} 
\caption{Clock period generated by the digitized timing module v.s. total read 0 delay.}\label{fig_tck}
\end{figure}
In Fig. \ref{fig_tck}, the clock period ($T_{CK}$) produced by the digitized timing module is displayed at different supply voltages. The clock period increases exponentially as V\textsubscript{DD} scales down. The digitized timing module, when compared to traditional methods, accurately tracks the read time of the bit cell in the array when the V\textsubscript{DD} enters the near-/sub-threshold domain (<0.5V). It eliminates the need for additional delay compensation when SRAM is operated in a wide voltage range. In other words, this technique prevents the introduction of any additional delay deviation and enhances PVT tracking ability. 

\subsection{Performance Comparison}

\begin{table}[tb!]\scriptsize
\begin{center}
\caption{Performance Comparison.}\label{tab2}
\resizebox{\linewidth}{!}{
\begin{tabular}{l|l|l|l|l}
\hline
& \cite{chien20180}    & \cite{do20160}    & \cite{do2019energy}  & Ultra8T     \\ \hline
Technology  & 28nmFDSOI    & 65nm        & 65nm        & 28nm          \\
Transistor  & 10T          & 8T          & 8T          & 8T            \\
V\textsubscript{DDMIN}      & 0.25V        & 0.2V        & 0.36V       & 0.25V         \\
Read Delay & 3.3\textmu s (0.25V) & 2.5\textmu s (0.2V) & 4\textmu s (0.36V)  & 1.11\textmu s (0.25V) \\
Min Energy  & 2.5pJ (0.3V)  & 1pJ (0.4V)   & 0.3pJ (0.5V) & 1.69pJ (0.4V)  \\ \hline
\end{tabular}}
\end{center}
\end{table}

Our scheme does not compromise read access speed even under sub-threshold voltage, and it achieves a minimum read access energy of 1.69 pJ at 0.4V. In comparison to recent studies listed in Table \ref{tab2}, the 10T cell structure used in operation \cite{chien20180} reduces the lowest energy consumption point to an even lower voltage, reaching the lowest reading energy point at 0.3V. However, due to the compensation method, the bit line discharge process in \cite{do20160} and \cite{do2019energy} is slower than that of traditional 8T SRAM bit line discharge. For example, the read access delay in \cite{do2019energy} is 4\textmu s at 0.36V. Designs \cite{do20160, do2019energy} using the 65nm process have lower leakage power, as a result, they have a lower energy consumption than Ultra8T SRAM. 

\section{Conclusion}
To improve the energy efficiency of low-power systems, scaling the supply voltage of SRAMs aggressively can significantly reduce their active and leakage power. 
%However, in sub/near-threshold operations, the relatively large leakage current impedes reliable read and write operations, constraining the lowest possible voltage of the system. 
In this paper, we present the Ultra8T SRAM with a leakage detection strategy to reduce the possible operating voltage aggressively while ensuring safe read access. 
%We first construct an analytical model to describe the relationship between leakage current, column depth, and the ratio of stored '0’s and '1’s in the column. By determining the function between leakage current and operation frequency, a safe working frequency at different V\textsubscript{DD} can be calculated. 
%A leakage detection strategy is implemented, where the voltage swing on a bit-line caused by leakage is converted into timing pulses by the proposed single-ended offset-canceling sensing amplifier circuits. A PVT-tracking variation-suppressed digitized timing scheme is also proposed to enhance the accurate leakage detection of Ultra8T. 
%Based on the quantified values of leakage current, Ultra8T can determine the minimum operating voltage and the safe frequency. 
We validate Ultra8T using a 256 rows $\times$ 64 columns array in 28nm CMOS technology, and post-simulations demonstrate that it can operate from the nominal voltage (0.8V) down to the sub-threshold voltage (0.25V).

\printcredits

\bibliographystyle{cas-model2-names}
\bibliography{refs}

\end{document}